\documentstyle[12pt]{article}
\newcommand{\nabf}{\mbox{\boldmath $\nabla$}}
\newcommand{\rhobf}{\mbox{\boldmath $\rho$}}

\newcommand{\Drm}{{\rm D}}
\newcommand{\pa}{\partial}

\newcommand{\text}{\rm}

\newcommand{\drm}{{\rm d}}

\newcommand{\Ccal}{{\cal C}}
\newcommand{\ug}{ \; = \; }
\newcommand{\ugg}{ \ = \ }

\newcommand{\Lra}{\Longrightarrow}

\newcommand{\ra}{\rightarrow}

\newcommand{\bb}{\begin{equation}}
\newcommand{\ee}{\end{equation}}
\newcommand{\bega}{\begin{eqnarray}}
\newcommand{\ega}{\end{eqnarray}}
\newcommand{\begae}{\begin{eqnarray*}}
\newcommand{\egae}{\end{eqnarray*}}

\newcommand{\h}{\hspace*{4ex}}
\newcommand{\dis}{\displaystyle}

\newcommand{\th}{\theta}
\newcommand{\Om}{\Omega}
\newcommand{\om}{\omega}

\newcommand{\cent}{\centerline}
\newcommand{\vs}{\vspace*}

\begin{document}

\baselineskip 0.65cm

\begin{center}

{\large {\bf Localized Superluminal Solutions to Maxwell Equations
propagating along a normal-sized waveguide}$^{\: (\dag)}$}
\footnotetext{$^{\: (\dag)}$  Work partially supported by CAPES
(Brazil), and by INFN, MURST and CNR (Italy).}

\end{center}

\vs{5mm}

\cent{ Michel Zamboni Rached }

\vs{0.5 cm}

\centerline{{\em Dep.to de F'isica, Universidade Estadual de Campinas, 
SP,
Brazil.}}

\vs{0.3 cm}

\centerline{\rm and}

\vs{0.3 cm}

\cent{ Erasmo Recami }

\vs{0.5 cm}

\cent{{\em Facolt\`a di Ingegneria, Universit\`a Statale di Bergamo, 
Dalmine
(BG), Italy;}}
\cent{{\em INFN---Sezione di Milano, Milan, Italy; \ and}}
\cent{{\em DMO--FEEC and CCS, State University of Campinas,
Campinas, S.P., Brazil.}}

\vs{1. cm}

\

\

{\bf Abstract  \ --} \ We show that localized (non-evanescent) solutions
to Maxwell equations exist, which propagate without distortion along
normal waveguides with Superluminal speed.\\

\


      
\newpage

{\bf 1. -- Introduction: Localized solutions to the wave equations}\\

\h Since 1915 Bateman[1] showed that Maxwell equations admit (besides of
the ordinary planewave solutions, endowed in vacuum with speed $c$) of
wavelet-type solutions, endowed in vacuum with group-velocities $0 \le v
\le c$. But Bateman's work went practically unnoticed. Only few authors, 
as Barut et al.[2] followed such a research line; incidentally, Barut et 
al. constructed even a wavelet-type solution travelling with Superluminal
group-velocity[3] $v > c$.

\h In recent times, however, many authors discussed the fact that all
(homogeneous) wave equations admit solutions with $0 < v < \infty$: \ see, 
e.g., Donnelly \& Ziolkowski[4], Esposito[4], Vaz \& Rodrigues[4]. Most of 
those authors confined themselves to investigate (sub- or Super-luminal)
{\em localized} non-dispersive solutions in vacuum: namely, those 
solutions
that were called ``undistorted progressive waves" by Courant \& Hilbert.
Among localized solutions, the most interesting appeared to be the 
so-called
``X-shaped" waves, which ---predicted even by Special Relativity in its
extended version[5]--- had been mathematically constructed by Lu \&
Greenleaf[6] for acoustic waves, and by Ziolkowski et al.[7], and later
Recami[8], for electromagnetism.

\h Let us recall that such ``X-shaped" localized solutions are Superluminal
(i.e., travel with $v > c$ in the vacuum) in the electromagnetic case;  and
are ``super-sonic" (i.e., travel with a speed larger than the sound-speed
in the medium) in the acoustic case. The first authors to produce 
X-shaped waves {\em experimentally} were Lu \& Greenleaf[9] for acoustics,
and Saari et al.[10] for optics.

\h Notwithstanding all that work, still it is not yet well understood 
what solutions (let us now confine ourselves to Maxwell equations and to
electromagnetic waves) have to enter into the play in many experiments.\\

\

{\bf 2. -- About evanescent waves}\\

\h Most of the experimental results, actually, did not refer to the
abovementioned localized, sub- or Super-luminal, solutions, which in 
vacuum are expected to propagate rigidly (or almost rigidly, when suitably
truncated).  The experiments most after fashion are, on the contrary,
those measuring the group-velocity of {\em evanescent waves\/}[cf., e.g.,
refs.11,12]. \ In fact, both Quantum Mechanics[13] and Special 
Relativity[5] had predicted tunnelling wavepackets (tunnelling photons too)
and/or evanescent waves to be Superluminal.

\h For instance, experiments[12] with evanescent waves travelling down 
an undersized {\em waveguide\/} revealed that evanescent modes are endowed
with Superluminal group-velocities[14].

\h A problem arises in connection with the experiment[15] with two
``barriers" 1 and 2 (i.e., segments of {\em undersized} waveguide). \ In
fact, it has been found that {\em for suitable frequency bands} the wave
coming out from barrier 1 goes on with practically infinite speed, 
crossing the intermediate {\em normal-sized} waveguide 3 in zero time. \
Even if
this can be theoretically understood by looking at the relevant transfer
function (see the computer simulations, based on Maxwell equations only, 
in refs.[16,17]), it is natural to wonder {\em what are} the solutions of 
Maxwell equations that can travel with Superluminal speed in a {\em normal} 
waveguide (where one normally meets ordinary propagating ---and not 
evanescent--- modes)...

\h Namely, the dispersion relation in undersized guides is \ $\om^2 - 
k^2 =
- \Om^2$, \ so that the standard formula \ $v \simeq \drm \om / \drm k$ 
\ yields a $v > c$ group-velocity[17,18]. \ However, in normal guides the 
dispersion
relation becomes \ $\om^2 - k^2 = + \Om^2$, \ so that the same formula 
yields values $v < c$ only.

\h We are going to show that actually localized solutions to Maxwell 
equations
propagating with $v > c$ do exist even in normal waveguides; but their
group-velocity $v$ {\em cannot} be given$^{\# 1}$ by the approximate
\footnotetext{$^{\# 1}$ Let us recall that the group-velocity is well 
defined only when the pulse has a clear bump in space; but it can be 
calculated
by the approximate, elementary relation \ $v \simeq \drm \om / \drm k$ \
{\em only} when some extra conditions are satisfied (namely, when $\om$ 
as a function of $k$ is also clearly bumped).}
formula \ $v \simeq \drm \om / \drm k$. \ One of the main motivations of 
the present note is just contributing to the clarification of this 
question.\\

\

{\bf 3. -- About some localized solutions to Maxwell equations.}\\

\h Let us start by considering localized solutions to Maxwell equations 
in vacuum. A {\em theorem} by Lu et al.[19] showed how to start from a 
solution
holding in the {\em plane} $(x,y)$ for constructing a threedimensional 
solution rigidly moving along the $z$-axis with Superluminal velocity $v$. \ 
Namely,
let us assume that $\psi (\rhobf; t)$, with $\rhobf \equiv (x,y)$, is a
solution of the 2-dimensional homogeneous wave equation:

\

\hfill{$
\left(\pa_x^2 + \pa_y^2 - {1 \over c^2} \pa_t^2 \right) \ \psi (\rhobf; 
t)
\ugg 0 \ .
$\hfill} (1)

\

By applying the {\em transformation} \ $\rhobf \ra \rhobf \; \sin\th$; \ 
\
$t \ra t - ({\cos\th} / c) \, z$, \  the angle $\th$ being fixed, with  $0 <
\th < \pi / 2$, one gets[19] that \ $\psi(\rhobf \; \sin\th ; \ t -
({\cos\th} / c) \, z)$ \ is a solution to the threedimensional homogeneous
wave-equation

\

\hfill{$
\left( \nabf^2 - {1 \over {c^2}} \pa_t^2 \right) \ \psi\!\left( \rhobf 
\;
\sin\th ; \ t - {\dis{\cos\th} \over  c} \, z \right) = 0 \ .
$\hfill} (2)

\

\h The mentioned theorem holds for the vacuum case, and in general is not 
valid when introducing boundary conditions. However we discovered that, in 
the case of a bidimensional solution $\psi$ valid on a circular domain of 
the $(x, y)$ plane, such that $\psi = 0$ for $|\rhobf| = 0$, the 
transformation above leads us to a (three-dimensional) {\em localized} 
solution rigidly travelling with Superluminal speed $v = c / \cos\theta$
inside a {\em cylindrical waveguide}; \ even if the waveguide radius 
$r$ will be no longer $a$, but $r = a / \sin\th > a$. \
We can therefore obtain an undistorted Superluminal solution propagating
down cylindrical (metallic) {\em waveguides} for each  (2-dimensional)
solution valid on a circular domain. \ Let us recall that, as well-known, 
any solution to the scalar wave equation corresponds to solutions of the
(vectorial) Maxwell equations (cf., e.g., ref.[8] and refs. therein).

\h For simplicity, let us put the origin O at the center of the circular
domain $\Ccal$, and choose a 2-dimensional solution that be axially
symmetric \ $\psi(\rho; t)$, \ with $\rho = |\rhobf|$, and with the 
initial
conditions \ $\psi(\rho; t=0) \ug \phi(\rho)$, \ and \ $\pa\psi / \pa 
t \ug
\xi(\rho)$ at $t = 0$.

\h Notice that, because of the transformations

\

\hfill{$
\rho \Lra \rho \; \sin\th
$\hfill} (3a)

\

\hfill{$
t \Lra t \; - \; {\dis{{\cos\th} \over c}} \, z \ ,
$\hfill} (3b)

\

the more the initial $\psi(\rho;t)$ is localized at $t=0$, the more the
(threedimensional) wave \ $\psi (\rho \; \sin\th; \  t - ({\cos\th} / c) z$  \
will be localized around $z = v t$. \ It should be also emphasized that,
because of transformation (3b), the velocity $c$ goes into the velocity
$v = c/cos\theta > c$.

\ \ Let us start with the formal choice

\

\hfill{$
\phi(\rho) \ugg {\dis{{\delta(\rho)} \over  \rho}} \ ; \ \ \ \ \ \ 
\xi(\rho)
\equiv 0 \ .
$\hfill} (4)

\

In cylindrical coordinates the wave equation (1) becomes 

\

\hfill{$
\left(\dis{{1 \over \rho}} \, \pa_\rho \rho \pa_\rho \; - \;
\dis{{1 \over {c^2}}} \, \pa_t^2 \right) \ \psi (\rho; t) \ugg 0 \ ,
$\hfill} (1')

\

which exhibits the assumed axial symmetry.  Looking for factorized 
solutions of the type \ $\psi(\rho; t) \ug R(\rho) \cdot T(t)$, \ one gets 
the equations \ $\pa_t^2 T = -\om^2 T$ \ and \ $(\rho^{-1} \pa_\rho + 
\pa_\rho^2 + \om^2/c^2)R = 0$, \ where the ``separation constant" $\om$ is 
a real parameter, which yield the solutions

\

\hfill{$
T \ugg A \, \cos\om t \; + \; B \, \sin\om t
$\hfill}

\hfill{  \hfill} (5)

\hfill{$
R \ugg C \; J_0(\dis{{\om \over c}} \, \rho) \ ,
$\hfill}

\

where quantities $A,B,C$ are real constants, and $J_0$ is the ordinary
zero-order Bessel function (we disregarded the analogous solution
$Y_0(\om \rho /c)$ since it diverges for $\rho = 0$). Finally, by 
imposing
the boundary condition $\psi = 0$ at $\rho = a$, one arrives at the 
base
solutions

\

\hfill{$
\psi(\rho;t) \ugg J_0(\dis{{k_n \over a}} \rho) \; \left( A_n \,
\cos\om_n t \; + \; B_n \, \sin\om_n t \right) \ ; \ \ \ \ k \equiv
\dis{{\om \over c}} a \ ,
$\hfill} (6)

\

the roots of the Bessel function being

$$k_n = {{\om_n a} \over c} \ .$$

\

\h The general solution for our bidimensional problem (with our boundary
conditions) will therefore be the Fourier-type series

\

\hfill{$
\Psi_{2\Drm}(\rho;t) \ugg \sum_{n=1}^{\infty} J_0(\dis{{k_n \over a}} \rho) \ 
\left(A_n \, \cos\om_n t \; + \;
B_n \, \sin\om_n t \right) \ .
$\hfill} (7)

\

The initial conditions (4) imply that \ $\sum A_n J_0(k_n \rho / a) \ug
\delta(\rho) / \rho$, \ and \ $\sum B_n J_0(k_n \rho / a) \ug 0$, \ so 
that all $B_n$ must vanish, while $A_n \ug 2 [a^2  J_1^2(k_n)]^{-1}$; \
and eventually one gets:

\

\hfill{$
\Psi_{2\Drm}(\rho;t) \ugg \sum_{n=1}^{\infty} \left( \dis{{{2} \over
{a^2 J_1^2(k_n)}}} \right) \; J_0(\dis{{k_n \over a}} \rho) \;
\cos\om_n t \ .
$\hfill} (8)\,\, ,

\

where $\om_n = k_n c / a$.

\h Let us explicitly notice that we can pass from such a formal solution 
to more physical ones, just by considering a finite number $N$ of terms. In 
fact, each partial expansion will satisfy (besides the boundary condition)
the second initial condition $\pa_t \psi = 0$ for $t= 0$, while the 
first initial
condition gets the form $\phi(\rho) = f(\rho)$, where $f(\rho)$ 
will be a (well) localized function, but no longer a delta-type function.
Actually, the ``localization" of $\phi(\rho)$ increases with increasing
$N$. \ We shall come back to this point below.

\

{\bf 4. -- Localized waves propagating Superluminally down 
(normal-sized) waveguides.}\\

\h We have now to apply transformations (3) to solution (8), in order to
pass to threedimensional waves propagating along a cylindrical (metallic)
waveguide with radius $r = a / \sin\th$. \ We obtain that Maxwell 
equations admit in such a case the solutions

\

\hfill{$
\Psi_{3\Drm}(\rho,z;t) \ugg \sum_{n=1}^{\infty} \left( \dis{{{2} \over
{a^2 J_1^2(k_n)}}} \right) \; J_0(\dis{{k_n \over a}} \rho \sin\th) \;
\cos\left[{{k_n \cos\theta} \over a} \; (z - \dis{{{c} \over {\cos\th}}} t) \right]
$\hfill} (9)

\

where \ $\om_n = k_n c / a$, \ which are sums over different propagating
modes.

\h Such solutions propagate, down the waveguide, rigidly with 
Superluminal
velocity$^{\# 2}$ \ $v = c / {\cos\th}$. \ Therefore, (non-evanescent) 
\footnotetext{$^{\# 2}$ Let us stress that each eq.(9) represents a
{\em multimodal} (but {\em localized}) propagation, as if the geometric
dispersion compensated for the multimodal dispersion.}
solutions to Maxwell equations exist, that are waves propagating 
undistorted
along {\em normal} waveguides with Superluminal speed (even if in
normal-sized waveguides the dispersion relation for each mode, i.e. for
each term of the Fourier-Bessel expansion, is the ordinary ``subluminal"
one, \ $\om^2/c^2 - k_z^2 = +\Om^2)$.

\h It is interesting that our Superluminal solutions travel rigidly down 
the waveguide: this is at variance with what happens for truncated 
(Superluminal) solutions[7-10], which travel almost rigidly only along their
finite ``field depth" and then abruptly decay.

\h Finally, let us consider a finite number of terms in eq.(8), at $t=0$. \
We made a few numerical evaluations: let us consider the results for $N=22$
(however, similar results can be already obtained, e.g., for $N=10$). \ The
first initial condition of eq.(4), then, is no longer a delta function, but
results to be the (bumped) bidimensional wave represented in Fig.1.

\h The threedimensional wave, eq.(9), corresponding to it, i.e., with the same
finite number $N = 22$ of terms, is depicted in Fig.2. \ It is still an
exact solution of the wave equation, for a metallic (normal-sized) waveguide  
with radius $r=a/\sin\theta$, propagating rigidly with Superluminal
group-velocity $v = c/\cos\theta$; \ moreover, it is now a {\em physical}
solution. \ In Fig.2 one can see its central portion, while in Fig.3 it is
shown the space profile along $z$, for $t = {\rm const.}$, of such a
propagating wave.

\

{\bf Acknowledgements} -- The authors are grateful to Flavio Fontana
(Pirelli Cavi, Italy) for having suggested the problem, and to Hugo E.
Hern\'andez-Figueroa (Fac. of Electric Engineering, UNICAMP) and Amr
Shaarawi (Cairo University) for continuous
scientific collaboration. \ Thanks are also due to Ant\^onio Chaves Maia Neto
for his kind help in the numerical evaluations, and to Franco Bassani, Carlo
Becchi, Rodolfo Bonifacio, Ray Chiao, Gianni Degli Antoni, Roberto Garavaglia,
Gershon Kurizki, Giuseppe Marchesini, Marcello Pignanelli, Andrea Salanti,
Abraham Steinberg and Jacobus Swart for stimulating discussions.

\vfill 
\newpage

\centerline{{\bf Figure Captions}}

\

{\bf Fig.1 --} Shape of the bidimensional solution of the wave equation
valid on the circular domain $\rho \le a; \ a = 0.1 \; {\rm mm}$ of the
$(x,y)$ plane, for $t = 0$, corresponding to the sum of $N = 22$ terms in the 
expansion (8). \ It is no longer a delta function, but it is still very
well peaked. \ By choosing it as the initial condition, instead of the first
one of eqs.(4), one gets the threedimensional wave depicted in Figs.2 and 3. \
The normalization condition is such that $|\Psi_{2\Drm}(\rho=0; \; t=0)|^2 
\ug 1$.

\

{\bf Fig.2 --} The (very well localized) threedimensional wave corresponding 
to the initial, bidimensional choice in Fig.1. \ It propagates rigidly (along 
the normal-sized circular waveguide with radius $r = a / \sin\th$) with
Superluminal speed $v = c / \cos\theta$. \ Quantity $\eta$ is defined as
$\eta \; \equiv \; (z - {{c} \over {\cos\th}} t)$. \ The normalization condition
is such that $|\Psi_{3\Drm}(\rho=0; \; \eta=0)|^2 \ug 1$. \

\

{\bf Fig.3 --} The shape along $z$, at $t=0$, of the threedimensional wave
whose main peak is shown in Fig.2.

\vfill
\newpage

\centerline{{\bf References}}

\

[1] H.Bateman: {\em Electrical and Optical Wave
Motion}  (Cambridge Univ.Press; Cambridge, 1915).\hfill\break

[2] A.O.Barut and H.C.Chandola:  {\em Phys. Lett.}
A180 (1993) 5. \ See also A.O.Barut: {\em Phys. Lett.} A189 (1994) 
277, and A.O.Barut et al.: refs.[3].\hfill\break

[3] A.O.Barut and A.Grant:  {\em Found. Phys. Lett.}
3 (1990) 303;  A.O.Barut and A.J.Bracken: {\em Found. Phys.}  22 (1992)
1267. \ See also refs.[14,19,20] below.\hfill\break

[4]  R.Donnelly and R.W.Ziolkowski: {\em Proc. Royal Soc. London}  A440
(1993) 541 [cf. also I.M.Besieris, A.M.Shaarawi and R.W.Ziolkowski:
{\em J. Math. Phys.} 30 (1989) 1254]; \ S.Esposito: {\em Phys. 
Lett} A225 (1997) 203; \ W.A.Rodrigues Jr. and J.Vaz Jr., {\em Adv. Appl.
Cliff. Alg.} S-7 (1997) 457.

[5] See, e.g., E.Recami: ``Classical tachyons and possible applications,"
{\em Rivista Nuovo Cimento}  9 (1986), issue no.6, pp.1-178; and
refs. therein.\hfill\break

[6] Jian-yu Lu and J.F.Greenleaf: {\em IEEE Transactions
on Ultrasonics, Ferroelectrics, and Frequency Control}
39 (1992) 19.\hfill\break

[7] R.W.Ziolkowski, I.M.Besieris and A.M.Shaarawi: {\em J. Opt. Soc. Am.} 
A10 (1993) 75.\hfill\break

[8] E.Recami: ``On localized `X-shaped' Superluminal solutions to
Maxwell equations", {\it Physica A} 252 (1998) 586.\hfill\break

[9] Jian-yu Lu and J.F.Greenleaf: {\em IEEE Transactions on
Ultrasonics, Ferroelectrics, and Frequency Control}  39
(1992) 441.\hfill\break

[10] P.Saari and K.Reivelt: ``Evidence of X-shaped propagation-invariant
localized light waves", {\em Phys. Rev. Lett.} 79 (1997) 4135. \ 
See also H.S\~{o}najalg, M.R\"{a}tsep and P.Saari: {\em Opt. Lett.} 22 
(1997) 310; \ {\em Laser Phys.} 7 (1997) 32).\hfill\break

[11] A.M.Steinberg, P.G.Kwiat and R.Y.Chiao: {\em Phys. Rev. Lett.}
71 (1993) 708, and refs. therein; \ {\em Scient. Am.} 269 (1993) issue no.2,
p.38. \ Cf. also R.Y.Chiao, A.E.Kozhekin, G.Kurizki: {\em Phys. Rev. Lett.}
77 (1996) 1254; \ {\em Phys. Rev.} A53 (1996) 586.\hfill\break

[12] A.Enders and G.Nimtz: {\em J. de Physique-I} {\bf 2} (1992) 1693; \
{\bf 3} (1993) 1089; \ {\bf 4} (1994) 1; \ H.M.Brodowsky, W.Heitmann and
G.Nimtz: J. de Physique-I {\bf 4} (1994) 565; {\em Phys. Lett. A}{\bf 222}
(1996) 125; {\em Phys. Lett. A\/}{\bf 196} (1994) 154; \ G.Nimtz and
W.Heitmann: {Prog. Quant. Electr.} {\em 21} (1997) 81.\hfill\break

[13] See V.S.Olkhovsky and E.Recami: {\em Phys. Reports} 214 (1992) 339,
and refs. therein; V.S.Olkhovsky et al.: {\em J. de Physique-I} 5 (1995)
1351-1365; \ T.E.Hartman: {\em J. Appl. Phys.} {\bf 33} (1962) 3427.\hfill\break

[14] Cf. A.P.L.Barbero, H.E.Hern\'andez-Figueroa and E.Recami: ``On the
propagation speed of evanescent modes" [LANL Archives \# 
physics/9811001],
submitted for pub., and refs. therein. \ Cf. also E.Recami, H.E.Hern\'andez F.,
and A.P.L.Barbero: {\em Ann. der Phys.} {\bf 7} (1998) 764.\hfill\break

[15] G.Nimtz, A.Enders and H.Spieker: {\em J. de Physique-I} 4 (1994) 565; \
``Photonic tunnelling experiments: Superluminal tunnelling", in {\it Wave
and Particle in Light and Matter -- Proceedings of the Trani Workshop,
Italy, Sept.1992}, ed. by A.van der Merwe and A.Garuccio (Plenum; New York,
1993).\hfill\break

[16] H.M.Brodowsky, W.Heitmann and G.Nimtz: {\em Phys. Lett.} A222 (1996)
125.\hfill\break

[17] R.Garavaglia: Thesis work (Dip. Sc. Informazione,
Universit\`a statale di Milano; Milan, 1998; G.Degli Antoni and
E.Recami supervisors).\hfill\break

[18] E.Recami and F.Fontana: ``Special Relativity and Superluminal
motions", submitted for publication.\hfill\break

[19] J.-y.Lu, H.-h.Zou and J.F.Greenleaf: {\em IEEE Transactions on 
Ultrasonics, Ferroelectrics and Frequency Control} 42 (1995) 850-853.\hfill\break

\end{document}